\documentclass[prd,twocolumn,showpacs,amsmath,amssymb,floatfix]{revtex4}
\usepackage{graphicx,color,dcolumn,booktabs,bm}
\usepackage{longtable,lscape}
\usepackage{txfonts}
\usepackage{amssymb}
\usepackage{indentfirst}

\begin{document}
\title{Predicted charged charmonium-like structures in the hidden-charm dipion decay of higher charmonia}
\author{Dian-Yong Chen$^{1,3}$}
\author{Xiang Liu$^{1,2}$\footnote{Corresponding author}}\email{xiangliu@lzu.edu.cn}
\affiliation{
$^1$Research Center for Hadron and CSR Physics,
Lanzhou University and Institute of Modern Physics of CAS, Lanzhou 730000, China\\
$^2$School of Physical Science and Technology, Lanzhou University, Lanzhou 730000,  China\\
$^3$Nuclear Theory Group, Institute of Modern Physics of CAS, Lanzhou 730000, China}
\date{\today}

\begin{abstract}

In this work, we predict two charged charmonium-like enhancement structures close to the $D^\ast\bar{D}$ and $D^\ast\bar{D}^\ast$ thresholds, where the Initial Single Pion Emission mechanism is introduced in the hidden-charm dipion decays of higher charmonia $\psi(4040)$, $\psi(4160)$, $\psi(4415)$ and charmonium-like state $Y(4260)$. We suggest BESIII to search for these structures in the $J/\psi\pi^+$, $\psi(2S)\pi^+$ and $h_b(1P)\pi^+$ invariant mass spectra of the $\psi(4040)$ decays into $J/\psi\pi^+\pi^-$, $\psi(2S)\pi^+\pi^-$ and $h_b(1P)\pi^+\pi^-$. In addition, the experimental search for these enhancement structures in the $J/\psi\pi^+$, $\psi(2S)\pi^+$ and $h_c(1P)\pi^+$ invariant mass spectra of the $\psi(4260)$ hidden-charm dipion decays will be accessible at Belle and BaBar.
\end{abstract}

\pacs{13.25.Gv, 14.40.Pq, 13.75.Lb} \maketitle

\section{introduction}

In the past years, experimentalist has made big progress on the
search for the charmonium-like states, the so-called XYZ states, in
the $B$ meson decay, the $e^+e^-$ collision, the $\gamma\gamma$
fusion process, which have aroused extensive
interest in revealing the underlying properties of the observed
charmonium-like states (see Refs.
\cite{Swanson:2006st,Zhu:2007wz,Godfrey:2008nc,Nielsen:2009uh,Brambilla:2010cs}
for a review). The study of charmonium-like states is a research
field full of challenges and opportunities in hadron physics.

Very recently the Belle Collaboration \cite{Collaboration:2011gj}
reported two charged $Z_b$ structures around 10610 MeV and 10650 MeV
by studying the $\Upsilon(nS)\pi^+$ ($n=1,2,3$) and
$h_b(mP)\pi^+$ ($m=1,2$) invariant mass spectra of $\Upsilon(5S)\to
\Upsilon(nS)\pi^+\pi^-,\,h_b(mP)\pi^+\pi^-$ hidden-bottom decay
channels (see Refs.
\cite{Liu:2008fh,Liu:2008tn,Bondar:2011ev,Chen:2011zv,Zhang:2011jj,Yang:2011rp,
Bugg:2011jr,russian,Guo:2011gu,Sun:2011uh,Chen:2011pv} for
theoretical progress). In Ref. \cite{Chen:2011pv}, we proposed the
Initial Single Pion Emission (ISPE) mechanism to explain the
observed $Z_b$ structures. By emitting a pion, $\Upsilon(5S)$ decays
into $B^{(*)}$ and $\bar{B}^{(*)}$ mesons with low momentum. Then,
$B^{(*)}$ and $\bar{B}^{(*)}$ mesons interact with each other by
exchanging $B^{(*)}$ meson and transit into $\Upsilon(nS)\pi^+$ or
$h_b(mP)\pi^+$. Here, two
structures near the $B{B}^*$ and $B^*B^*$ thresholds appear in the
$\Upsilon(nS)\pi^+$ and $h_b(mP)\pi^+$ invariant mass spectra, which
could correspond to $Z_b(10610)$ and $Z_b(10650)$
\cite{Collaboration:2011gj}.

Just indicated in Ref. \cite{Chen:2011pv}, if the ISPE mechanism is
an universal mechanism existing in heavy quarkonium decay, we can
naturally extend such physical picture to study hidden-charm decays
of higher vector charmonia due to the similarity between charmonium and
bottomonium families, and predict some novel phenomena similar to
the $Z_b$ structures.

In Particle Data Book \cite{Nakamura:2010zzi}, six vector charmonia
are established well, which are $J/\psi$, $\psi(2S)$, $\psi(3770)$,
$\psi(4040)$, $\psi(4160)$ and $\psi(4415)$ with $I^G(J^{PC})=0^-(1^{--})$. Among these charmonia,
only $\psi(4040)$, $\psi(4160)$ and $\psi(4415)$ are higher than the
thresholds of $D\bar{D}$, $D\bar{D}^*$ and $D^*\bar{D}^*$.
Thus, we
study the hidden-charm decays of $\psi(4040)$, $\psi(4160)$ and $\psi(4415)$ via the ISPE mechanism.
From the analysis of such
modes, one indicates that enhancement structures similar to the charged $Z_b$ also exist in the charm case.

In addition, in this work we will study the hidden-charm decays of $Y(4260)$,
which is an important charmonium-like state observed by the BaBar
Collaboration in the $e^+e^-\to \gamma_{ISR}J/\psi\pi^+\pi^-$
process \cite{Aubert:2005rm}. Its mass, width and $J^{PC}$ are
$4263^{+8}_{-9}$ MeV, $95\pm{14}$ MeV and $1^{--}$
\cite{Nakamura:2010zzi}. The study presented in Ref.
\cite{Chen:2010nv} indicates that $Y(4260)$ can be related to
charmonia $\psi(4160)$ and $\psi(4415)$, where the $Y(4260)$
structure can be reproduced by the interference of production
amplitudes of the $e^+e^-\to J/\psi\pi^+\pi^-$ processes via
direct $e^+e^-$ annihilation and through intermediate charmonia
$\psi(4160)$ and $\psi(4415)$ \cite{Chen:2010nv}. We naturally apply the ISPE mechanism existing in $\psi(4160)$ and $\psi(4415)$ decays to
discuss the hidden-charm dipion decays of $Y(4260)$. Thus, studying
$Y(4260)$ hidden-charm decays through the ISPE mechanism is an
intriguing issue, where we will also predict some enhancement structure similar to the charged $Z_b$. As announced by BaBar \cite{Aubert:2005rm}, $Y(4260)$ was first observed in its $J/\psi\pi^+\pi^-$ decay channel. Searching enhancement structure in the $J/\psi\pi$ invariant mass spectrum of $Y(4260)\to J/\psi\pi^+\pi^-$ will be accessible in future experiments, which could provide a direct test to the non-resonant explanation for $Y(4260)$ proposed in \cite{Chen:2010nv}.

This work is organized as follows. After the
Introduction, we illustrate the hidden-charm dipion decays of higher charmonia under the ISPE mechanisms. In Sec. III, the numerical results are presented. The last section is the discussion and conclusion.

\section{The hidden-charm decays of higher charmonia}

\subsection{The ISPE mechanism}

With $\psi(4040)\to J/\psi\pi^+\pi^-$ as an example, we first illustrate the possible decay mechanisms in
the dipion hidden-charm decay of higher charmonium. $\psi(4040)$ can directly decay into $J/\psi\pi^+\pi^-$.
In Refs. \cite{Kuang:1981se,Yan:1980uh, Novikov:1980fa}, the QCD Multipole Expansion method
was proposed and can be applied to calculate such direct decay process. The second mechanism is that the dipion is from the intermediate scalar ($\sigma(600), f_0(980)$) or tensor ($f_2(1270)$) meson, where the hadronic loops constructed by the $D^{(*)}$ mesons could be as a bridge to connect $\psi(4040)$ and $J/\psi\pi^+\pi^-$ (see Ref. \cite{Chen:2011qx} for more details).

\begin{center}
\begin{figure}[htb]
\begin{tabular}{cccc}
\scalebox{0.48}{\includegraphics{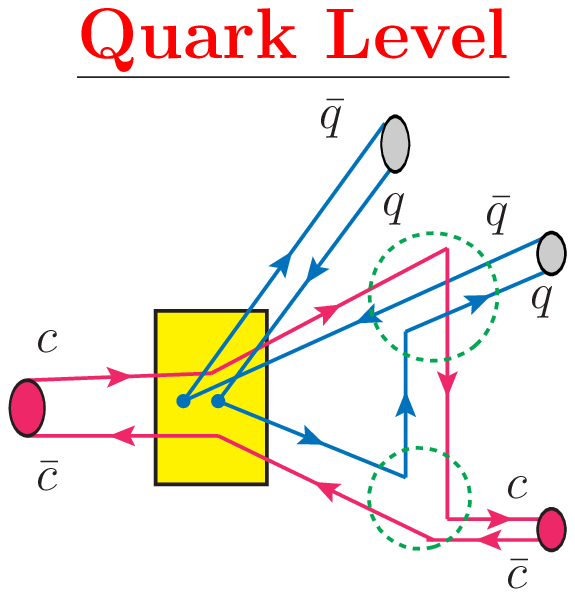}}&\raisebox{3.5em}{\large{$\Longleftrightarrow$}}&
\scalebox{0.7}{\includegraphics{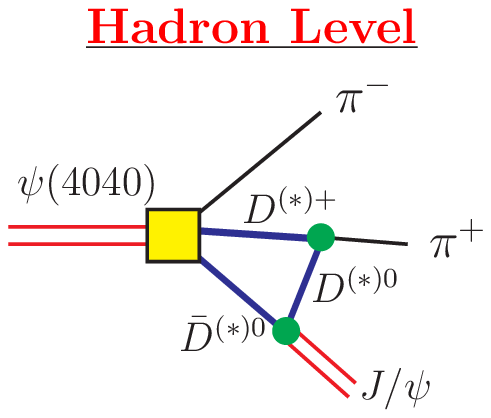}}&\raisebox{3.5em}{\large{$+\cdots$}}
\end{tabular}
\caption{(Color online.) The quark-level (left-side diagram) and hadron-level (right-side diagram) descriptions of the ISPE mechanism existing in the hidden-charm decays of higher charmonia.  \label{IPE}}
\end{figure}
\end{center}

The remaining decay mechanism existing in the hidden-charm dipion decays of higher charmonia is the ISPE mechanism, which was first proposed in Ref. \cite{Chen:2011pv}. By the quark-level diagram we give an explicit description (left-side diagram in Fig. \ref{IPE}) of the ISPE mechanism in $\psi(4040)\to J/\psi\pi^+\pi^-$ decay. The physical picture is that with a pion emission $\psi(4040)$ first dissolves into $D^{(*)}$ and $\bar{D}^{(*)}$ mesons with low momentum, which further turn into $J/\psi\pi^+$. Here, $D^{(*)}\bar{D}^{(*)}\to J/\psi\pi^+$ transition occurs via exchanging $D^{(*)}$ meson \cite{Chen:2011pv}.

An equivalent hadron-level description is also presented in the right-side diagram of Fig.
\ref{IPE}, which can be as an effective approach for dealing
with the practical calculations.

\subsection{Effective Lagrangian and coupling constant}\label{h}

We adopt effective Lagrangian approach to calculate these hadron-level diagrams listed in Fig. \ref{IPE}. Here, the effective Lagrangians involved
in the interaction vertexes in Fig. \ref{IPE} include \cite{Oh:2000qr,Casalbuoni:1996pg,Colangelo:2002mj}
\begin{eqnarray}
&&\mathcal{L}_{\psi^\prime D^{(*)} D^{(*)} \pi}\nonumber\\&&=-ig_{\psi^\prime DD
\pi} \varepsilon^{\mu \nu \alpha \beta} \psi^\prime_{\mu} \partial_{\nu} D
\partial_{\alpha} \pi \partial_{\beta} \bar{D} + g_{\psi^\prime D^\ast D \pi} {\psi^\prime}^{\mu} (D \pi
\bar{D}^\ast_{\mu} + D^\ast_{\mu} \pi \bar{D}) \nonumber\\
&&\quad-ig_{\psi^\prime D^\ast D^\ast \pi} \varepsilon^{\mu \nu \alpha
\beta} \psi_{\mu}^\prime D^\ast_{\nu} \partial_{\alpha} \pi
\bar{D}^\ast_\beta -ih_{\psi^\prime D^\ast D^\ast \pi} \varepsilon^{\mu \nu
\alpha \beta} \partial_{\mu} \psi_{\nu}^\prime D^\ast_{\alpha} \pi
\bar{D}^\ast_{\beta},\nonumber
\end{eqnarray}
where $\psi^\prime$ denotes the initial state charmonium (one of the $\psi(4040)$, $\psi(4160)$, $\psi(4415)$ or $Y(4260)$). This Lagrangian reflects
the initial state charmonium decays into
$D^{(*)} \bar{D}^{(*)} \pi$.
\begin{eqnarray}
&&\mathcal{L}_{D^\ast D^{(\ast)} \pi} \nonumber\\&&= ig_{D^\ast D
\pi} (D^\ast_{\mu} \partial^\mu \pi \bar{D}-D \partial^\mu \pi
\bar{D}^\ast_{\mu})-g_{D^\ast D^\ast \pi} \varepsilon^{\mu \nu
\alpha \beta}
\partial_{\mu} D^\ast_{\nu} \pi \partial_{\alpha}
\bar{D}^\ast_{\beta},\nonumber
\end{eqnarray}
\begin{eqnarray}
&&\mathcal{L}_{\psi D^{(*)} D^{(*)}}\nonumber\\&&=ig_{\psi DD}
\psi_{\mu} (\partial^\mu D \bar{D}- D \partial^\mu \bar{D})-g_{\psi
D^\ast D} \varepsilon^{\mu \nu \alpha \beta}
\partial_{\mu} \psi_{\nu} (\partial_{\alpha} D^\ast_{\beta} \bar{D}
\nonumber\\&&\quad+ D \partial_{\alpha}
\bar{D}^\ast_{\beta})-ig_{\psi D^\ast D^\ast} \big\{ \psi^\mu
(\partial_{\mu} D^{\ast \nu} \bar{D}^\ast_{\nu} -D^{\ast \nu}
\partial_{\mu}
\bar{D}^\ast_{\nu}) \nonumber\\
&&\quad+ (\partial_{\mu} \psi_{\nu} D^{\ast \nu} -\psi_{\nu}
\partial_{\mu} D^{\ast \nu}) \bar{D}^{\ast \mu} + D^{\ast \mu}(\psi^\nu \partial_{\mu} \bar{D}^\ast_{\nu} -
\partial_{\mu} \psi^\nu \bar{D}^\ast_{\nu})\big\},\nonumber
\end{eqnarray}
\begin{eqnarray}
&&\mathcal{L}_{h_c D^{(*)} D^{(*)}}\nonumber\\&&= g_{h_c D^\ast D}
h_c^\mu (\bar{D}^\ast_{\mu} D + D^\ast_\mu \bar{D})+ ig_{h_c D^\ast
D^\ast} \varepsilon^{\mu \nu \alpha \beta}
\partial_{\mu} h_{c \nu} D^\ast_{\alpha} \bar{D}^\ast_{\beta}, \nonumber
\end{eqnarray}
which will be applied to describe a rescattering mechanism involving in the two charmed
mesons into $J/\psi \pi$ or $h_c \pi$ by exchanging a $D^{(*)}$ meson. In the above Lagrangians,
$D$ and $D^*$ are grouped together on the basis of heavy quark symmetry while that pions appear as the result of a
representation of the chiral symmetry. In addition, we define charm meson iso-doublets as $D^{(*)}=(D^{(*)0},D^{(*)+})$,
${\bar{D}^{(*)T}}=(\bar{D}^{(*)0},D^{(*)-})$ and $\pi={\mbox{\boldmath $\tau$}}\cdot {\mbox{\boldmath $\pi$}}$ \cite{Oh:2000qr}.

The values of the coupling constants can be determined by the relations
\begin{eqnarray}
g_{\psi DD} &=& g_{\psi D^\ast D^\ast} \frac{m_D}{m_D^\ast} =g_{\psi
D^\ast D} m_{\psi} \sqrt{\frac{m_D}{m_D^\ast}}
=\frac{m_{\psi}}{f_{\psi}}, \nonumber\\
g_{h_c DD^\ast} &=& -2g_1 \sqrt{m_{h_c} m_D m_{D^\ast}} , \ \ g_{h_c
D^\ast D^\ast} =2 g_1 \frac{m_{D^\ast}}{\sqrt{m_{h_c}}}, \nonumber\\
g_{D^\ast D^\ast \pi} &=&  \frac{g_{D^\ast D \pi}}{\sqrt{m_D
m_{D^\ast}}} =\frac{2 g}{f_\pi},\ \
g_1=-\sqrt{\frac{m_{\chi_{c0}}}{3}} \frac{1}{f_{\chi_{c0}}},\nonumber
\end{eqnarray}
where $f_{\psi}=0.416$ GeV and $f_{\chi_{c0}}=0.297$ GeV are the decay constants of
$\psi$ and $\chi_{c0}$, respectively. In addition, $f_{\chi_{c0}} \simeq 0.51 $ GeV can be approximately determined by the QCD sum rule approach \cite{Colangelo:2002mj}. With the measured branching ratio of
$D^\ast \to D \pi$ by CLEO-c \cite{Anastassov:2001cw} and $f_{\pi}=132$ MeV, one gets $g=0.59$ \cite{Isola:2003fh}.


\subsection{Decay Amplitudes}

With these Lagrangians just listed above, we write out the decay amplitude for the dipion
transition between $\psi(4040)$ and $J/\psi$, i.e., there are three interaction vertexes and three $D^{(*)}$ propagators, which are obtained by the effective Lagrangian presented in Sec. \ref{h}. Additionally, we also introduce the monopole form factor $\mathcal{F}(q^2)$ in decay amplitudes, which is taken as $\mathcal{F}(q^2)= (\Lambda^2-m_E^2)/(q^2-m_E^2)$. Here, $m_E$ is the
mass of the exchanged meson while the phenomenological
parameter $\Lambda$ can be parameterized as $\Lambda =m_E + \beta
\Lambda_{QCD}$ with $\Lambda_{QCD}=220$ MeV. Such monopole form factor is introduced to describe the
structure effects of the interaction vertexes as well as the off-shell
effects of the exchanged charmed mesons for $D^{(\ast)}\bar{D}^{(\ast)}\to J/\psi\pi^\pm, h_c(1P)\pi^\pm$ transitions in $\psi(4040)\to J/\psi\pi^+\pi^-, h_c(1P)\pi^+\pi^-$ decays.

\begin{center}
\begin{figure}[htb]
\begin{tabular}{cccc}
\scalebox{0.42}{\includegraphics{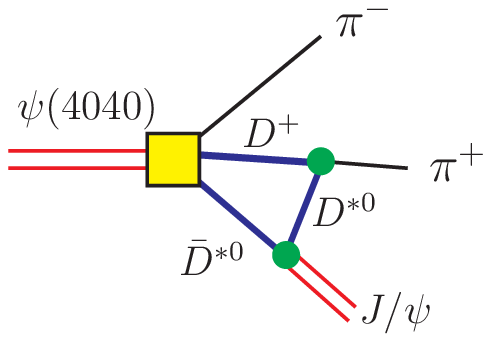}}&\scalebox{0.42}{\includegraphics{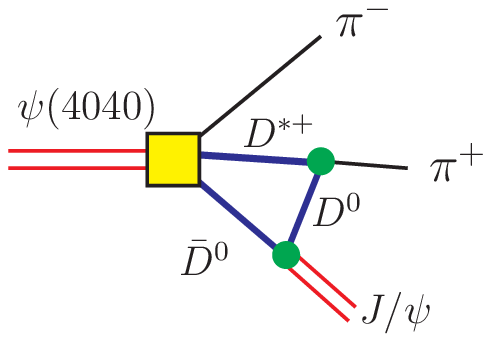}}&\scalebox{0.42}{\includegraphics{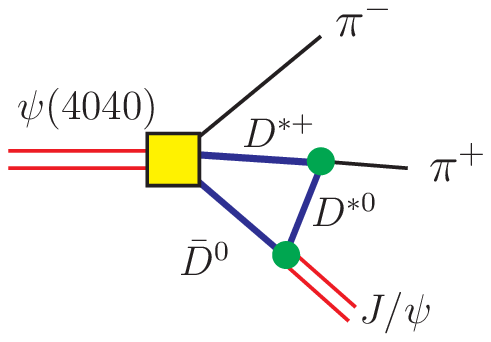}}&
\scalebox{0.42}{\includegraphics{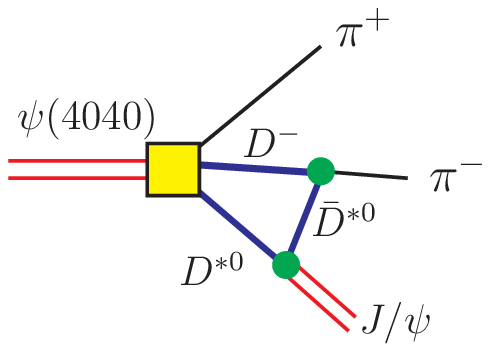}}\\
(1)&(2)&(3)&(4)\\\scalebox{0.42}{\includegraphics{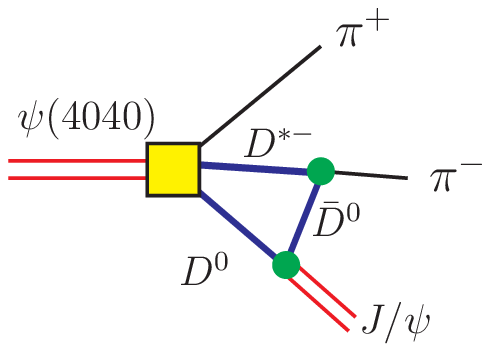}}&\scalebox{0.42}{\includegraphics{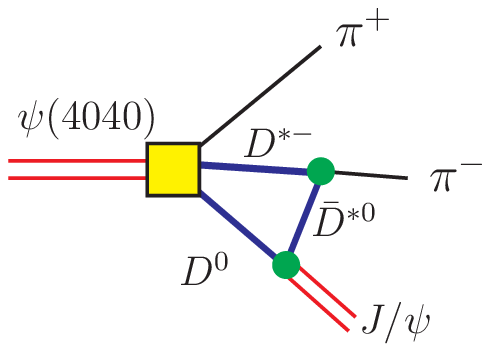}}&
\scalebox{0.42}{\includegraphics{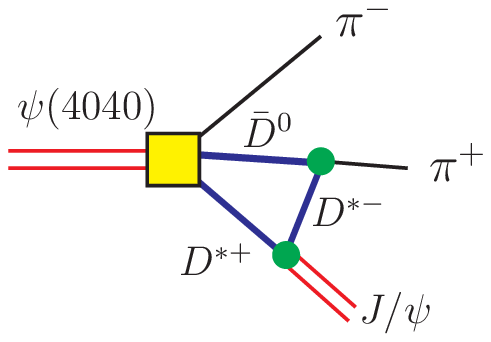}}&\scalebox{0.42}{\includegraphics{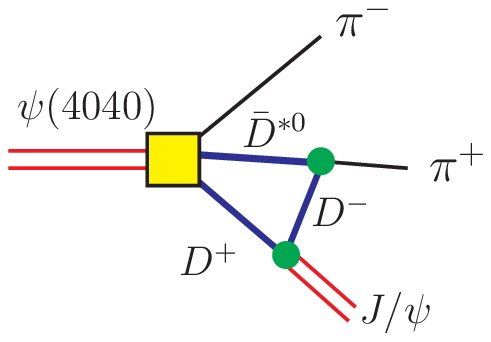}}\\
\\
(5)&(6)&(7)&(8)\\
\scalebox{0.42}{\includegraphics{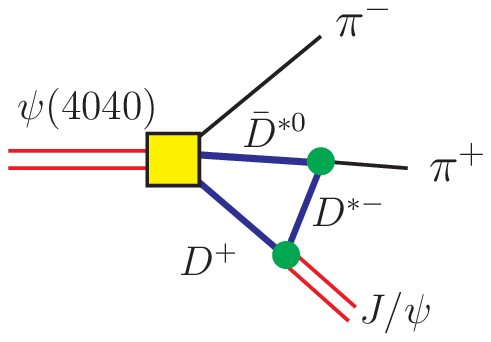}}&
\scalebox{0.42}{\includegraphics{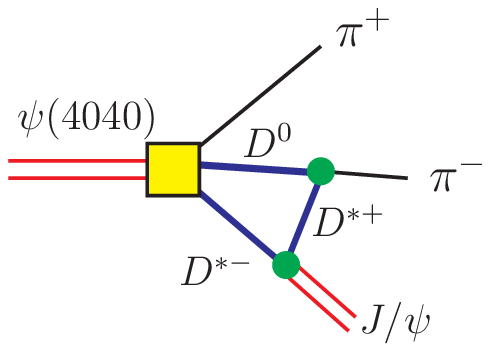}}&\scalebox{0.42}{\includegraphics{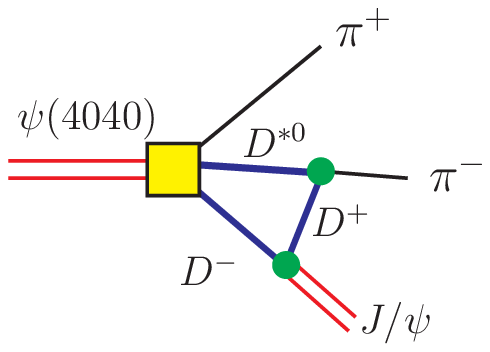}}&\scalebox{0.42}{\includegraphics{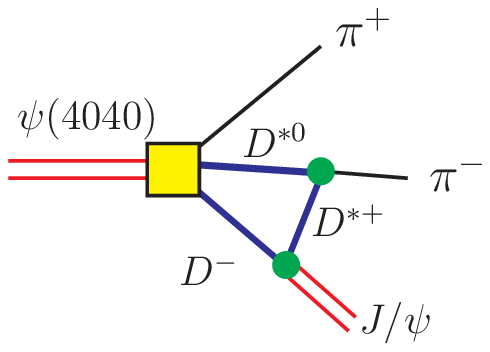}}\\
(9)&(10)&(11)&(12)
\end{tabular}
\caption{(Color online.) The hadron-level diagrams for $\psi(4040)\to J/\psi\pi^+\pi^-$ decays with
$D^\ast \bar{D} + h.c$ as the intermediate states.  \label{DDS}}
\end{figure}
\end{center}

When only considering the intermediate $D^\ast\bar{D}+h.c.$ contributions to $\psi(4040)\to J/\psi\pi^+\pi^-$,
there are twelve diagrams just shown in Fig. \ref{DDS}. Among these diagrams, there are only six independent
diagrams if considering $SU(2)$ symmetry, i.e., Fig. \ref{DDS} (i) can be transferred into Fig. \ref{DDS} (i+6) ($i=1,\cdots,6$) by transformations $D^{(*)+}\rightleftharpoons \bar{D}^{(*)0}$ and $D^{(*)-}\rightleftharpoons {D}^{(*)0}$. Thus, the total decay amplitude for $\psi(4040)\to J/\psi(p_5)\pi^+(p_3)\pi^-(p_4)$
with the intermediate $D^\ast(p_1)\bar{D}(p_2)+D(p_1)\bar{D}^\ast(p_2)$ contributions are expressed as
\begin{eqnarray}
\mathcal{M}[\psi(4040)\to J/\psi\pi^+\pi^-]_{D^\ast\bar{D}+h.c.}=2\sum_{i=1,\cdots,6}M_{D^\ast\bar{D}+h.c.}^{(i)},
\end{eqnarray}
where we mark the four momenta of the corresponding mesons. Factor 2 reflects $SU(2)$ symmetry mentioned above. The subscript $D^\ast\bar{D}+h.c.$ denotes that $\psi(4040)\to J/\psi\pi^+\pi^-$ occurs via the
intermediate $D^\ast\bar{D}+D\bar{D}^*$. The expressions of decay amplitudes $M_{D^\ast\bar{D}+h.c.}^{(i)}$ ($i=1,2,3$) read as
\begin{eqnarray}
&&M_{D^\ast\bar{D}+h.c.}^{(1)}  = (i)^3 \int
\frac{d^4 q}{(2\pi)^4} [g_{\psi^\prime D^\ast D \pi} \epsilon_{\psi}^\mu] [
i g_{D^\ast D^\ast \pi} (iP_4^\rho)]
\nonumber\\
&&\quad\times  [-ig_{J/\psi D^\ast D^\ast}
\epsilon_{J/\psi}^\nu((-iq_\nu+ip_{2 \nu})g_{\theta \phi} + (iP_{5
\phi}+ iq_{\phi}) g_{\nu \theta}  \nonumber\\
&&\quad -(ip_{2\theta}ip_{5\theta}) g_{\nu \phi})] \frac{1}{
p_1^2-m_D^2}\frac{-g_{\mu}^\phi +p_{1 \mu}
p_{1}^{\phi}/m_{D^\ast}^2}{p_2^2 - m_{D^\ast}^2 }
\nonumber\\
&& \quad  \times \frac{-g_{\rho}^{ \theta} +q_\rho
q^\theta/m_{D^\ast}^2}{q^2-m_{D^\ast}^2}
\mathcal{F}^2(q^2),\label{ha1}
\end{eqnarray}
\begin{eqnarray}
&&M_{D^\ast\bar{D}+h.c.}^{(2)} = (i)^3 \int
\frac{d^4 q}{(2\pi)^4} [g_{\psi^\prime D^\ast D \pi} \epsilon_{\psi}^\mu]
[ig_{D^\ast D \pi} (-ip_4^\rho)]
\nonumber\\
&&\quad \times [ig_{J/\psi DD} \epsilon_{J/\psi}^\nu
(ip_{2\nu}-iq_{\nu})] \frac{-g_{\mu \rho} +p_{1 \mu} p_{1
\rho}/m_{D^\ast}^2
}{p_1^2-m_{D^\ast}^2}\nonumber\\
&&\quad \times \frac{1}{p_2^2-m_D^2} \frac{1}{q^2-m_D^2}
\mathcal{F}^2(q^2),\label{ha2}
\end{eqnarray}
\begin{eqnarray}
&&M_{D^\ast\bar{D}+h.c.}^{(3)} = (i)^3 \int
\frac{d^4 q}{(2\pi)^4} [g_{\psi^\prime D^\ast D \pi} \epsilon_{\psi}^\mu]
[-g_{D^\ast D^\ast \pi} \varepsilon^{\theta
\phi \delta \tau} (iq^\theta) \nonumber\\
&&\quad \times (-ip_1^\delta)] [-g_{J/\psi D^\ast D}
\varepsilon^{\rho \nu
\alpha \beta} (ip_{5 \rho}) \epsilon_{J/\psi \nu} (-iq_{\alpha})]\nonumber\\
&&\quad \times \frac{-g_{\mu \tau}+p_{1 \mu} p_{1
\tau}/m_{D^\ast}^2}{p_1^2-m_{D^\ast}^2} \frac{1}{p_2^2-m_D^2}
\nonumber\\&&\quad\times\frac{-g_{\beta \phi}+q_\beta q_\phi/m_{D^\ast}^2}{q^2-
m_{D^\ast}^2} \mathcal{F}^2(q^2),\label{ha3}
\end{eqnarray}
which correspond to the dipion transitions between $\psi(4040)$ and $J/\psi$ with a initial single pion ($\pi^-$) emission. $M_{D^\ast\bar{D}+h.c.}^{(4)}$, $M_{D^\ast\bar{D}+h.c.}^{(5)}$ and $M_{D^\ast\bar{D}+h.c.}^{(6)}$
can be obtained by $M_{D^\ast\bar{D}+h.c.}^{(1)}$, $M_{D^\ast\bar{D}+h.c.}^{(2)}$ and $M_{D^\ast\bar{D}+h.c.}^{(3)}$ respectively if making the replacement $p_3\rightleftharpoons p_4$ in Eqs. (\ref{ha1})-(\ref{ha3}). Here, $M_{D^\ast\bar{D}+h.c.}^{(j)}$ ($j=4,5,6$) are decay amplitudes of
the dipion transitions between $\psi(4040)$ and $J/\psi$ with a initial single pion ($\pi^+$) emission.

\begin{center}
\begin{figure}[htb]
\begin{tabular}{cccc}
\scalebox{0.42}{\includegraphics{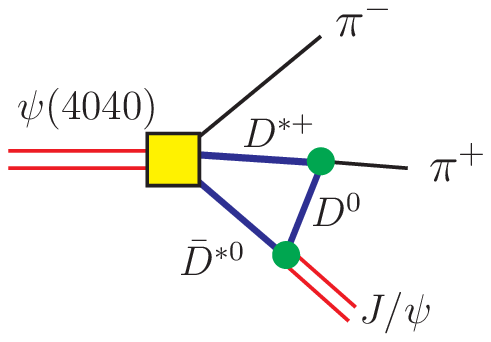}}&\scalebox{0.42}{\includegraphics{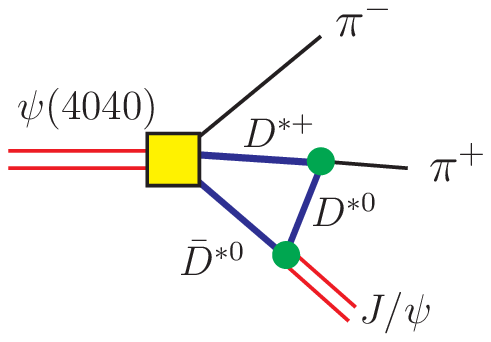}}
&\scalebox{0.42}{\includegraphics{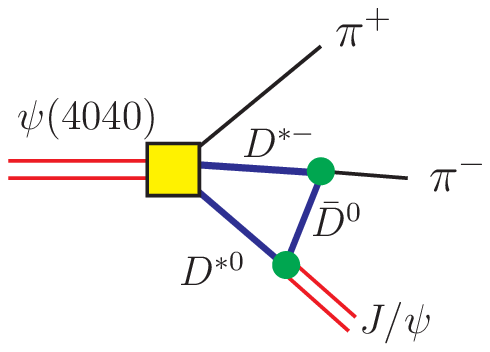}}&\scalebox{0.42}{\includegraphics{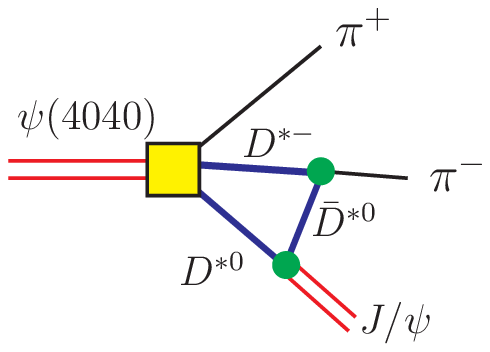}}\\
(1)&(2)&(3)&(4)\\
\scalebox{0.42}{\includegraphics{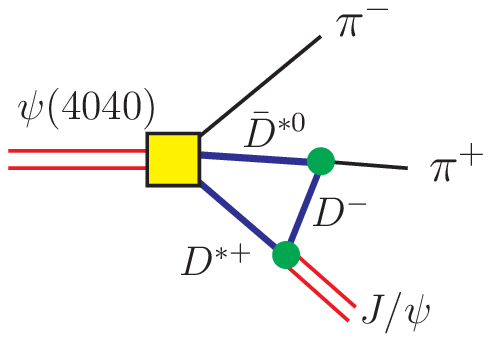}}&\scalebox{0.42}{\includegraphics{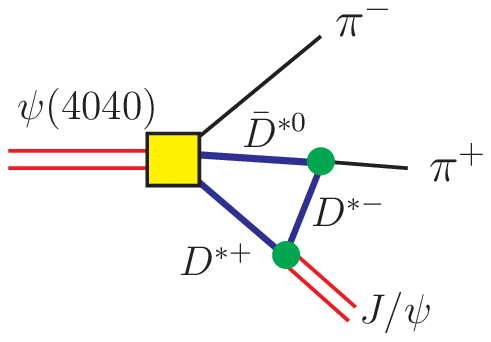}}
&\scalebox{0.42}{\includegraphics{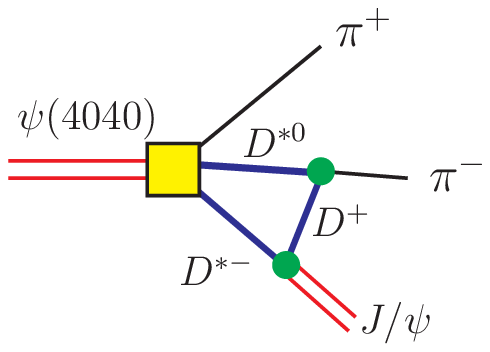}}&\scalebox{0.42}{\includegraphics{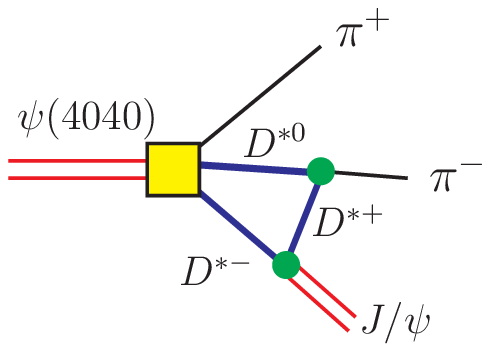}}\\
(5)&(6)&(7)&(8)
\end{tabular}
\caption{(Color online.) The hadron-level diagrams for $\psi(4040)\to J/\psi\pi^+\pi^-$ decays with
$D^\ast \bar{D}^\ast$ as the intermediate states.  \label{DSDS}}
\end{figure}
\end{center}

We also present the decay amplitude of $\psi(4040)\to J/\psi(p_5)\pi^+(p_3)\pi^-(p_4)$ via the
intermediate $D^\ast(p_1)\bar{D}^\ast(p_2)$.
\begin{eqnarray}
\mathcal{M}[\psi(4040)\to J/\psi\pi^+\pi^-]_{D^\ast\bar{D}^\ast}=2\sum_{\alpha=1,\cdots,4}M_{D^\ast\bar{D}^\ast}^{(\alpha)}.\label{a1}
\end{eqnarray}
We list all diagrams contributing to $\psi(4040)\to J/\psi\pi^+\pi^-$ in Fig. \ref{DSDS}. Among these eight
eight diagrams,  Fig. \ref{DDS} ($\alpha$) can be obtained by Fig. \ref{DDS} ($\alpha+4$) ($\alpha=1,\cdots,4$) if making the transformations $D^{(*)+}\rightleftharpoons \bar{D}^{(*)0}$ and $D^{(*)-}\rightleftharpoons {D}^{(*)0}$, which results in factor 2 in Eq. (\ref{a1}) due to $SU(2)$ symmetry.

The decay amplitudes $M_{D^\ast\bar{D}^\ast}^{(1)}$ and $M_{D^\ast\bar{D}^\ast}^{(2)}$ are expressed as
\begin{eqnarray}
&&M_{D^\ast\bar{D}^\ast}^{(1)} = (i)^3 \int
\frac{d^4 q}{(2\pi)^4} [-ig_{\psi^\prime D^\ast D^\ast \pi}
\varepsilon^{\mu \rho \alpha \beta} \epsilon_{\psi \mu} (ip_{3
\alpha})\nonumber\\
&&\quad -ih_{\psi^\prime D^\ast D^\ast \pi}
\varepsilon^{\alpha \mu \rho \beta} \epsilon_{\psi \mu}(-ip_{0
\alpha})][ig_{D^\ast D
\pi} (-ip_{4 \lambda})] \nonumber \\
&&\quad \times [-g_{J/\psi D^\ast D} \varepsilon_{\delta \nu \theta
\phi} (ip_{5}^{\delta}) \epsilon_{J/\psi}^{\nu} (-ip_{2}^{\theta})]
\frac{-g_{\rho}^{\lambda} +p_{1 \rho}
p_1^\lambda/m_{D^\ast}^2 }{p_1^2-m_{D^\ast}^2}\nonumber \\
&&\quad \times  \frac{-g_{\beta}^{ \phi} +p_{2 \beta}
p_2^\phi/m_{D^\ast}^2}{ p_2^2-m_{D^\ast}^2} \frac{1}{q^2-m_D^2}
\mathcal{F}^2(q^2),\label{hb1}
\end{eqnarray}
\begin{eqnarray}
&&M_{D^\ast\bar{D}^\ast}^{(2)} = (i)^3 \int
\frac{d^4 q}{(2\pi)^4} [-ig_{\psi^\prime D^\ast D^\ast \pi}
\varepsilon^{\mu \rho \alpha \beta} \epsilon_{\psi \mu} (ip_{3
\alpha})  \nonumber \\
&&\quad -ih_{\psi^\prime D^\ast D^\ast \pi} \varepsilon^{\alpha \mu \rho
\beta} \epsilon_{\psi \mu} (-ip_{0 \alpha})] [-g_{D^\ast D^\ast \pi}
\varepsilon^{\delta \tau \theta \phi} (-ip_{1 \delta})
(iq_{\theta})] \nonumber\\
&&\quad \times [-ig_{J/\psi D^\ast D^\ast} \epsilon_{J/\psi}^\nu (
(-iq_{\nu} +ip_{2\nu})) g_{\omega \lambda} + (ip_{5 \omega}
+iq_{\omega})g_{\nu \lambda} \nonumber\\
&&\quad +(-ip_{2\lambda} -ip_{5 \lambda}) g_{\nu
\omega}]\frac{-g^{\rho \tau} +p_{1\rho}
p_{1\tau}/m_{D^\ast}^2}{p_1^2-m_{D^\ast}^2}
\nonumber\\
&&\quad \times  \frac{-g_{\beta}^{ \omega} + p_{2\beta}
p_2^\omega/m_{D^\ast}^2 }{p_2^2-m_{D^\ast}^2}
\frac{-g_{\phi}^{\lambda} +q_\phi q^\lambda/m_{D^\ast}^2
}{q^2-m_{D^\ast}^2} \mathcal{F}^2(q^2).\label{hb2}
\end{eqnarray}
Thus, by Eqs. (\ref{hb1}) and (\ref{hb2}) we can easily obtain decay amplitudes $M_{D^\ast\bar{D}^\ast}^{(3)}$ and $M_{D^\ast\bar{D}^\ast}^{(4)}$ corresponding to Fig. \ref{DSDS} (3) and (4), where
the transformation $p_3\rightleftharpoons p_4$ is performed.

In the following, we extend the same framework to study the dipion transition between $\psi(4040)$ and $h_c(1P)$. By replacing $J/\psi$ with $h_c(1P)$ in Fig. \ref{DDS} (1), (3), (4), (6), (7), (9), (10) and (12) and Fig. \ref{DSDS}, we obtain all
diagrams relevant to $\psi(4040)\to h_c(1P)\pi^+\pi^-$ decay. The total decay amplitudes of $\psi(4040)\to h_c(1P)(p_5)\pi^+(p_3)\pi^-(p_4)$ via $D^\ast(p_1)\bar{D}(p_2)+D(p_1)\bar{D}^\ast(p_2)$ and $D^\ast(p_1)\bar{D}^\ast(p_2)$ are
\begin{eqnarray}
\mathcal{M}[\psi(4040)\to h_c(1P)\pi^+\pi^-]_{D^\ast\bar{D}+h.c.}&=&2\sum_{\beta=1,\cdots,4}A_{D^\ast\bar{D}+h.c.}^{(\beta)},\label{c1}\\
\mathcal{M}[\psi(4040)\to h_c(1P)\pi^+\pi^-]_{D^\ast\bar{D}^\ast}&=&2\sum_{\kappa=1,\cdots,4}A_{D^\ast\bar{D}^\ast}^{(\kappa)},\label{c2}
\end{eqnarray}
respectively, where the concrete amplitude expressions are
\begin{eqnarray}
&&A_{D^\ast\bar{D}+h.c.}^{(1)} =(i)^3 \int
\frac{d^4 q}{(2\pi)^4} [g_{\psi^\prime D^\ast D^\ast \pi}
\epsilon_{\psi}^\mu ] \nonumber\\
&&\quad \times [ig_{D^\ast D \pi} (-ip_4^\rho)] [ig_{h_c D^\ast
D^\ast} \varepsilon_{\delta \nu \theta \phi} (ip_5^\delta)
\epsilon_{h_c}^\nu]  \frac{1}{p_1^2-m_D^2}  \nonumber \\
&&\quad \times \frac{-g_{\mu}^{ \phi} +p_{2\mu}
p_2^\phi/m_{D^\ast}^2}{p_2^2 -m_{D^\ast}^2} \frac{-g_{\rho}^{
\theta}+ q_\rho q^\theta/m_{D^\ast}^2
}{q^2-m_{D^\ast}^2} \mathcal{F}^2(q^2), 
\end{eqnarray}
\begin{eqnarray}
&&A_{D^\ast\bar{D}+h.c.}^{(2)} = (i)^3
\int \frac{d^4 q}{(2\pi)^4} [g_{\psi^\prime D^\ast D^\ast \pi}
\epsilon_{\psi}^\mu ] [-g_{D^\ast D^\ast \pi} \varepsilon_{\theta
\phi \delta \tau} (iq^\theta)
\nonumber\\
&&\quad \times (-ip_1^\delta)] [g_{h_c D^\ast D} \epsilon_{h_c \nu}]
\frac{-g_{\mu}{ \phi} +p_{1 \mu} p_1^\phi/m_{D^\ast}^2}{
p_1^2-m_{D^\ast}^2}\nonumber\\
&&\quad \times \frac{1}{p_2^2-m_D^2} \frac{-g^{\nu \tau} +q^\nu
q^\tau/m_{D^\ast}^2}{q^2-m_{D^\ast}^2} \mathcal{F}^2(q^2),
\end{eqnarray}
and
\begin{eqnarray}
&&A_{D^\ast\bar{D}^\ast}^{(1)} = (i)^3
\int \frac{d^4 q}{(2\pi)^4} [-ig_{\psi^\prime D^\ast D^\ast \pi}
\varepsilon_{\mu \rho \alpha \beta} \epsilon_{\psi}^\mu (ip_3^\alpha
-ip_0^\alpha)]  \nonumber\\
&&\quad \times [ig_{D^\ast D \pi} (-ip_{4 \lambda})][g_{h_c D^\ast
D} \epsilon_{h_c \nu}] \frac{-g^{\rho \lambda}+ p_1^\rho
p_1^\lambda/m_{D^\ast}^2}{ p_1^2-m_{D^\ast}^2} \nonumber\\
&&\quad \times  \frac{-g^{\beta \nu}+p_2^\beta
p_2^\nu/m_{D^\ast}^2}{p_2^2-m_{D^\ast}^2}
\frac{1}{q^2-m_D^2}\mathcal{F}^2(q^2),
\end{eqnarray}
\begin{eqnarray}
&&A_{D^\ast\bar{D}^\ast}^{(2)} = (i)^3
\int \frac{d^4 q}{(2\pi)^4} [-ig_{\psi^\prime D^\ast D^\ast \pi}
\varepsilon_{\mu \rho \alpha \beta} \epsilon_{\psi}^\mu (ip_3^\alpha
-ip_0^\alpha)]  \nonumber\\
&&\quad \times[-g_{D^\ast D^\ast \pi} \epsilon_{\delta \tau \theta
\phi} (-ip_1^\delta) (iq^\theta)] [ig_{h_c D^\ast D^\ast}
\varepsilon_{\kappa \nu \lambda \omega} (ip_5^\kappa)
\epsilon_{h_c}^\nu]\nonumber\\
&&\quad \times \frac{-g^{\rho \tau} +p_1^\rho
p_1^\tau/m_{D^\ast}^2}{p_1^2-m_{D^\ast}^2} \frac{-g^{\beta
\omega}+p_2^\beta p_2^\omega/m_{D^\ast}^2 }{ p_2^2- m_{D^\ast}^2 }\nonumber\\
&&\quad \times \frac{-g^{\phi \lambda}+ q^\phi
q^\lambda/m_{D^\ast}^2}{q^2-m_{D^\ast}^2}\mathcal{F}^2(q^2).
\end{eqnarray}
After performing the transformation $p_3\rightleftharpoons p_4$,
$A_{D^\ast\bar{D}+h.c.}^{(1)}$, $A_{D^\ast\bar{D}+h.c.}^{(2)}$, $A_{D^\ast\bar{D}^\ast}^{(1)}$ and $A_{D^\ast\bar{D}^\ast}^{(2)}$
can be transferred into $A_{D^\ast\bar{D}+h.c.}^{(3)}$, $A_{D^\ast\bar{D}+h.c.}^{(4)}$,  $A_{D^\ast\bar{D}^\ast}^{(3)}$ and $A_{D^\ast\bar{D}^\ast}^{(4)}$ respectively.

The differential decay width for $\psi(4040)
\to J/\psi \pi^+ \pi^-$ reads as
\begin{eqnarray}
d\Gamma =\frac{1}{3} \frac{1}{(2 \pi)^3} \frac{1}{32
m_{\psi(4040)}^3} \overline{|\mathcal{M}^2|}
dm_{J/\psi \pi^+}^2 dm_{\pi^+\pi^-}^2
\end{eqnarray}
with $m_{J/\psi \pi^+}^2 = (p_4 + p_5)^2$ and $m_{\pi^+\pi^-}^2
=(p_3 +p_4)^2$, where the overline indicates the
average over the polarizations of the $\psi(4040)$ in the
initial state and the sum over the polarization of $J/\psi(4040)$ in
the final state. Replacing $m_{J/\psi\pi^+}$ with $m_{\psi(2S)\pi^+}$ or $m_{h_c(1P)\pi^+}$, we obtain
the differential decay width for $\psi(4040)
\to \psi(2S) \pi^+ \pi^-$ or $\psi(4040)
\to h_c(1P) \pi^+ \pi^-$.

When studying the hidden-charm dipion decay of other higher charmonia $\psi(4160)$, $\psi(4415)$ and charmonium-like state $Y(4260)$, we only need to replace the relevant coupling constants and the masses in the formulism of the $\psi(4040)$ decays.

\section{numerical result}

In this work, we are mainly concerned with the line shapes of the differential decay widths of $\psi(4040)$, $\psi(4160)$, $\psi(4415)$ and charmonium-like state $Y(4260)$ decays into
$J/\psi\pi^+\pi^-$, $\psi(2S)\pi^+\pi^-$ and $h_c(1P)\pi^+\pi^-$, which are dependent on the invariant mass spectra of $J/\psi\pi^+$, $\psi(2S)\pi^+$ and $h_c(1P)\pi^+$. Thus, we set the coupling constants of $\psi D^{(\ast)}\bar{D}^{(\ast)}\pi$ as 1 in our calculation. Besides these coupling constants listed in Sec.  \ref{h}, other input parameters are the masses involved in our calculation, which are taken from Particle Data Book \cite{Nakamura:2010zzi}.

\begin{figure*}[hbtp]
\begin{center}
\begin{tabular}{cc}
\LARGE{$\psi(4040)$}& \LARGE{$\psi(4160)$}\\
\includegraphics[height=130pt]{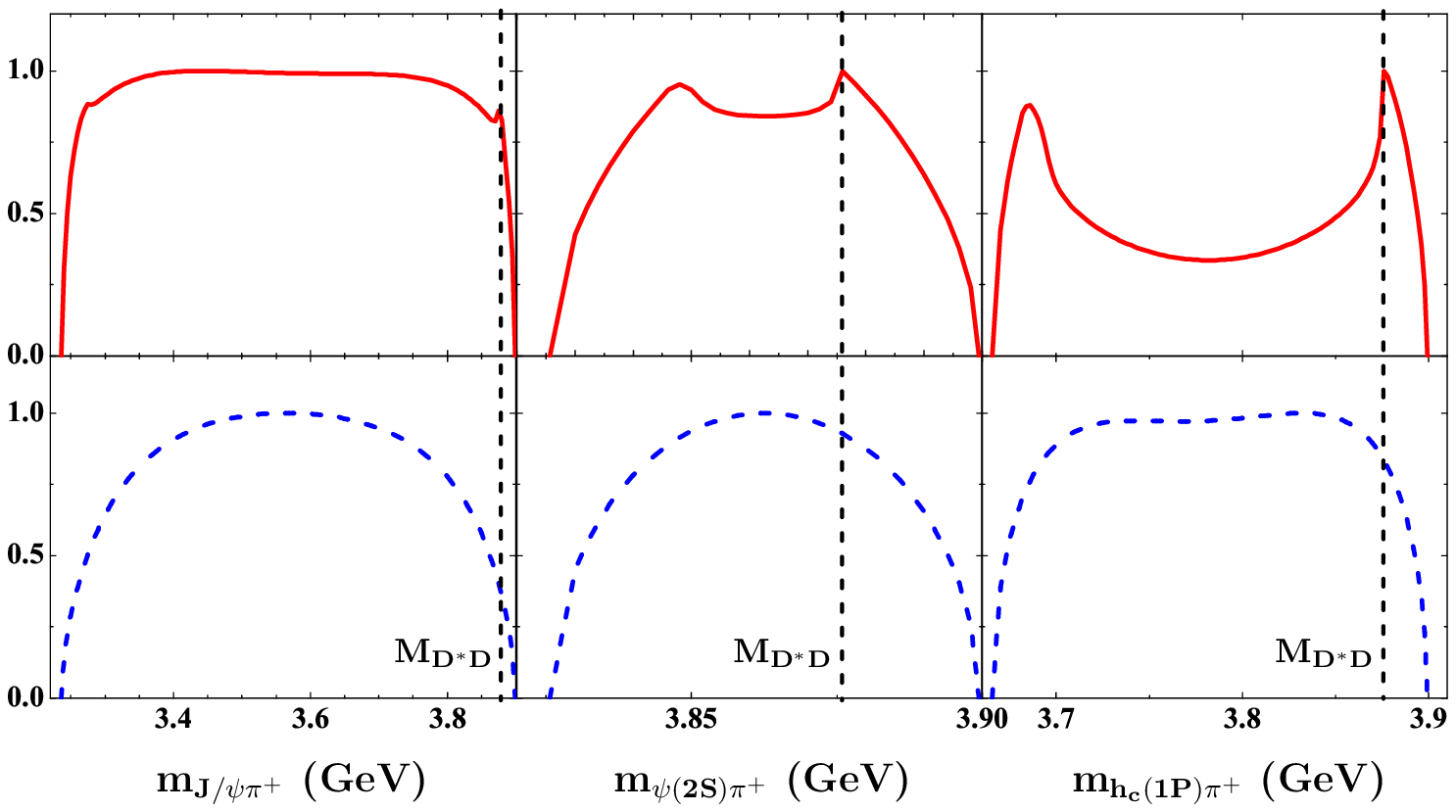}&
\hspace{5mm}
\includegraphics[height=130pt]{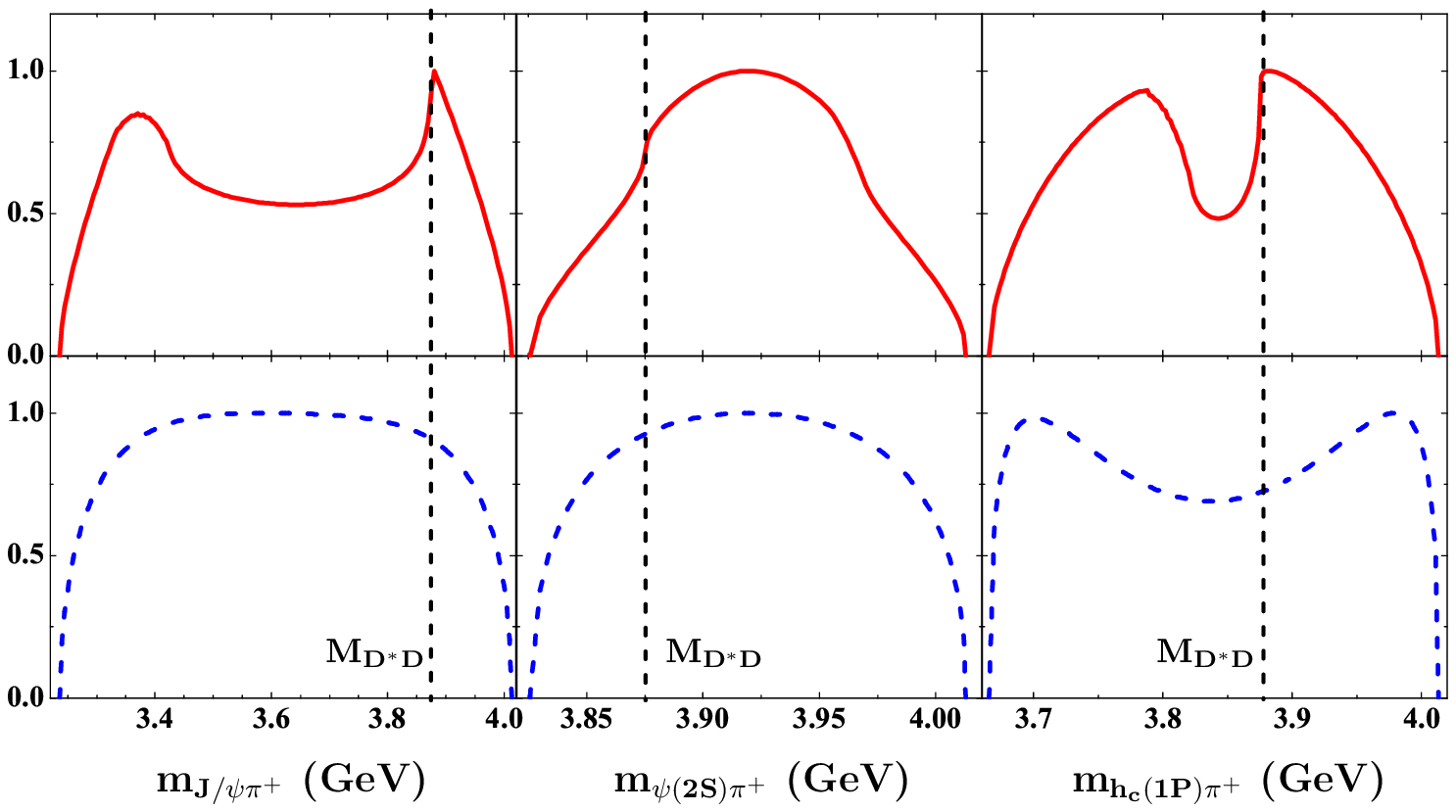}\\
\\
\LARGE{$\psi(4415)$}&\LARGE{$Y(4260)$}\\
\includegraphics[height=130pt]{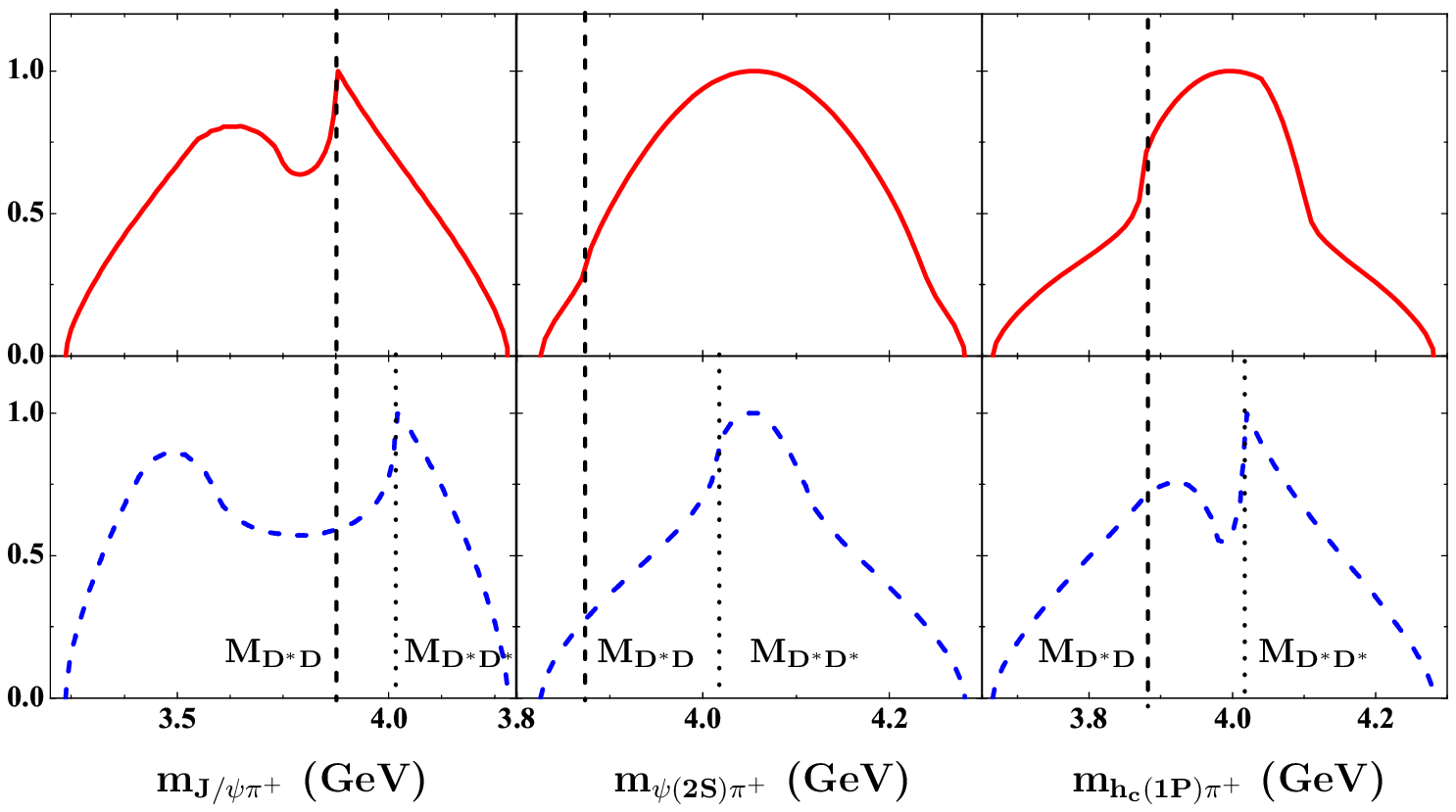}&
\hspace{5mm}
\includegraphics[height=130pt]{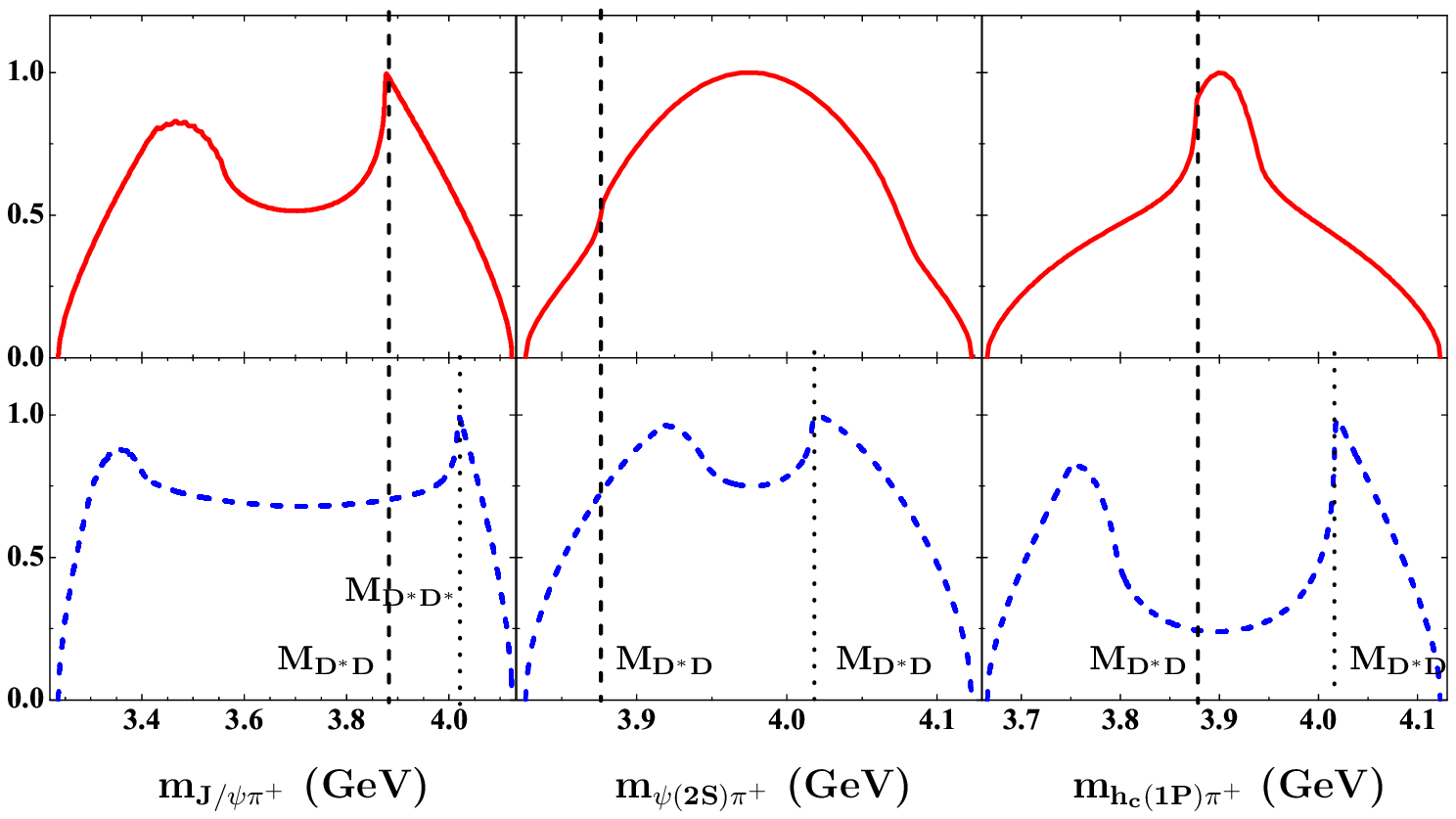}
\end{tabular}
\end{center}
\caption{(Color online.) The invariant mass spectra of
$J/\psi\pi^+$, $\psi(2S)\pi^+$ and $h_c(1P)\pi^+$ for the
$\psi(4040)$, $\psi(4160)$, $\psi(4415)$ and $Y(4260)$ decays into
$J/\psi\pi^+\pi^-$, $\psi(2S)\pi^+\pi^-$ and $h_c(1P)\pi^+\pi^-$.
Here, the solid, dashed correspond to the results considering
intermediate  $D\bar{D}^*+h.c.$ and $D^*\bar{D}^*$ respectively in
Fig. \ref{IPE}. The vertical dashed lines and the dotted lines
denote the threshhold of $D^\ast \bar{D}$ and $D^\ast
\bar{D}^\ast$ respectively. Here, the maximum of the line shape is normalized to 1.  \label{result}}
\end{figure*}

In Fig. \ref{result}, we present the results of $d\Gamma/d m_{J/\psi\pi^+}$, $d\Gamma/d m_{\psi(2S)\pi^+}$ and $d\Gamma/d m_{h_c(1P)\pi^+}$ of $\psi(4040)$, $\psi(4160)$, $\psi(4415)$, $Y(4260)$ decays into $J/\psi\pi^+\pi^-$, $\psi(2S)\pi^+\pi^-$, $h_c(1P)\pi^+\pi^-$.

\begin{enumerate}

\item{There exit sharp peak structures close to the $D^\ast \bar{D}$ threshold and the corresponding reflections in the distributions of $d\Gamma/d m_{J/\psi\pi^+}$, $d\Gamma/d m_{\psi(2S)\pi^+}$ and $d\Gamma/d m_{h_c(1P)\pi^+}$ of $\psi(4040)\to J/\psi\pi^+\pi^-$, $\psi(4040)\to \psi(2S)\pi^+\pi^-$ and $\psi(4040)\to h_c(1P)\pi^+\pi^-$ decays. We notice that this structure appearing in $d\Gamma(\psi(4040)\to J/\psi\pi^+\pi^-)/d m_{J/\psi\pi^+}$ is not obvious comparing with the structure in $d\Gamma(\psi(4040)\to \psi(2S)\pi^+\pi^-)/d m_{\psi(2S)\pi^+}$ or $d\Gamma(\psi(4040)\to h_c(1P)\pi^+\pi^-)/d m_{h_c(1P)\pi^+}$ distribution. }

\item{Two sharp peaks appear in the $d\Gamma(\psi(4160)\to J/\psi\pi^+\pi^-)/d m_{J/\psi\pi^+}$ and $d\Gamma(\psi(4160)\to h_c(1P)\pi^+\pi^-)/d m_{h_c(1P)\pi^+}$ distributions, which are close the $D^\ast \bar{D}$ threshold. The structure in the $J/\psi\pi^+$ invariant mass spectrum is more narrow than that in the $h_c(1P)\pi^+$ invariant mass spectrum. }

\item{In the hidden-charm dipion decays of $\psi(4415)$, we find two sharp peak structures around the $D^\ast \bar{D}$ and $D^\ast \bar{D}^\ast$ thresholds appearing in the $J/\psi\pi^+$ invariant mass spectra. In addition, a sharp peak close the $D^\ast \bar{D}^\ast$ threshold is observed in the $h_c(1P)\pi^+$ invariant mass spectrum distribution. In the $d\Gamma(\psi(4415)\to \psi(2S)\pi^+\pi^-)/d m_{\psi(2S)\pi^+}$ distribution, a peak near $D^\ast\bar{D}^\ast$ with its reflection form a broad structure. Under the ISPE mechanism, the intermediate $D^*\bar{D}$ can result in a very broad structure in the $h_c(1P)\pi^+$ invariant mass spectrum distribution.}

\item{There exist the sharp peaks close to $D^\ast \bar{D}$ threshold in the $d\Gamma(\psi(4260)\to J/\psi\pi^+\pi^-)/d m_{J/\psi\pi^+}$ and $d\Gamma(\psi(4260)\to h_c(1P)\pi^+\pi^-)/d m_{h_c(1P)\pi^+}$ distributions, the structures around $D^\ast \bar{D}^\ast$ threshold in the $d\Gamma(\psi(4260)\to \psi(2S)\pi^+\pi^-)/d m_{\psi(2S)\pi^+}$ and $d\Gamma(\psi(4260)\to h_c(1P)\pi^+\pi^-)/d m_{h_c(1P)\pi^+}$ distributions. The peak close the $D^\ast \bar{D}$ threshold and its reflection overlap with each each to form a broad structure in the $h_c\pi^+$ invariant mass spectrum. }

\end{enumerate}

\begin{figure}[hbtp]
\begin{center}
\scalebox{0.8}{\includegraphics{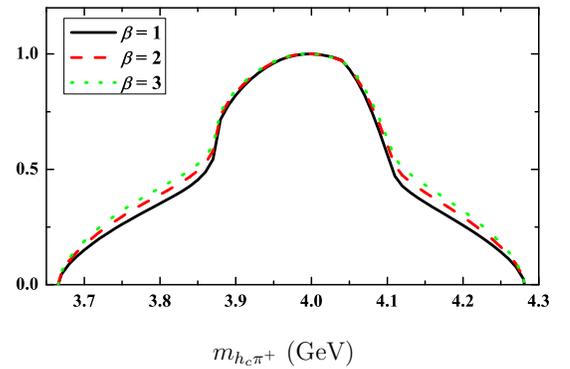}}
\end{center}
\caption{(Color online.) The dependence of $d\Gamma(\psi(4415)\to h_c(1P)\pi^+\pi^-)/d m_{h_c(1P)\pi^+}$ distribution on $\beta$. Here, $\psi(4415) \to h_c(1P) \pi^+ \pi^-$ occurs via the intermediate $D^\ast \bar{D} + h.c$.
\label{beta}}
\end{figure}

We need to specify that the result presented in Fig. \ref{result} is obtained by taking
$\beta=1$. Our study shows that the line shapes in Fig. \ref{result} are weakly dependent on
the values of $\beta$. With $\psi(4415)\to h_c\pi^+\pi^-$ as an example, in Fig. \ref{beta} we illustrate
the $\beta$ dependence of $d\Gamma(\psi(4415)\to h_c(1P)\pi^+\pi^-)/d m_{h_c(1P)\pi^+}$ distribution, where the line shapes corresponding to $\beta=1,2,3$ remain almost unchanged.

\section{discussion and conclusion}

In this work, we study the line shapes of the differential decay widths of $\psi(4040)$, $\psi(4160)$, $\psi(4415)$ and charmonium-like state $Y(4260)$ decays into $J/\psi\pi^+\pi^-$, $\psi(2S)\pi^+\pi^-$ and $h_c(1P)\pi^+\pi^-$, where the ISPE mechanism is introduced. Furthermore, we predict the sharp peak structures close to $D^\ast \bar{D}$ and $D^\ast \bar{D}^\ast$ thresholds appearing the corresponding $J/\psi\pi^+$, $\psi(2S)\pi^+$ and $h_c(1P)\pi^+$ invariant mass spectra.

The ISPE mechanism plays crucial role to form these novel charged charmonium-like structures in the hidden-charm dipion decays of higher charmonia. To some extent, these predicted structures are the charmonium analogue of two newly observed $Z_b$ structures in the hidden-bottom dipion decays of $\Upsilon(5S)$ \cite{Collaboration:2011gj}.

We suggest further experimental search for these predicted charmonium-like structures close to the $D^\ast \bar{D}$ and $D^\ast \bar{D}^\ast$ thresholds. Recently, BESIII has stated accumulating $\psi(4040)$ data with an aim to search for higher charmonia and the charmonium-like states \cite{bes}. Our result shows the charged structures around the $D^\ast \bar{D}$ threshold in the $J/\psi\pi^+$, $\psi(2S)\pi^+$ and $h_b(1P)\pi^+$ invariant mass spectra of $\psi(4040)$ decays into $J/\psi\pi^+\pi^-$, $\psi(2S)\pi^+\pi^-$ and $h_b(1P)\pi^+\pi^-$, which are accessible at BESIII and could be considered in future studies.

Since these charged charmonium-like structures also exist in the $J/\psi\pi^+$, $\psi(2S)\pi^+$ and $h_b(1P)\pi^+$ invariant mass spectra of $\psi(4260)$ hidden-charm dipion decays, carrying out the search for them will be an important and intriguing research topic, especially at Belle and BaBar.

If these predicted enhancement structures can be found in the higher charmonium and Y(4260) hidden-charm decays, it will provide direct test to the ISPE mechanism existing in the higher charmonium or Y(4260) hidden-charm dipion decays.

\begin{figure}[hbtp]
\begin{center}
\scalebox{0.8}{\includegraphics{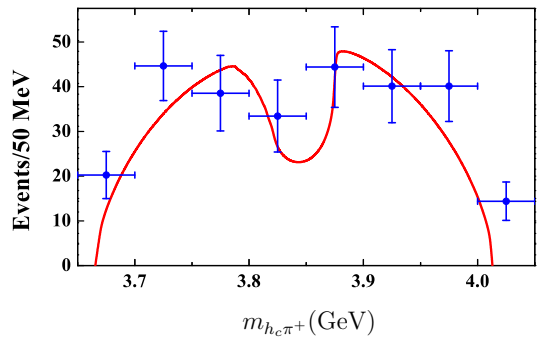}}
\end{center}
\caption{(Color online.) A comparison of the $h_c\pi^\pm$ mass distribution of $\psi(4160)\to h_c(1P) \pi^+\pi^-$ (red solid line) predicted in this work and measurement by CLEO-c (blue points with errors) \cite{:2011uqa}. Here, CLEO-c measured the $h_c(1P)\pi^\pm$ mass distribution from $e^+e^-\to h_c(1P) \pi^+\pi^-$ at $E_{CM}=4170$ MeV \cite{:2011uqa}. We
normalize our numbers for a real comparison with the available CLEO-c data.
\label{com}}
\end{figure}

Note added: after completion of this work, we notice a measurement reported by CLEO-c \cite{:2011uqa}. Recently the CLEO-c Collaboration announced the measurement of the $h_c(1P)\pi^\pm$ mass distribution from $e^+e^-\to h_c(1P) \pi^+\pi^-$ at $E_{CM}=4170$ MeV (points with errors in Fig. 4 (b) of Ref. \cite{:2011uqa}). Thus, we compare the predicted $h_c\pi^\pm$ mass distribution of $\psi(4160)\to h_c(1P) \pi^+\pi^-$ (see Fig. \ref{result}) with the CLEO-c result. We notice that the predicted theoretical line shape of the $h_c(1P)\pi^\pm$ mass distribution of $\psi(4160)\to h_c(1P) \pi^+\pi^-$ in this work is consistent with
the one measured by CLEO-c, which is listed in Fig. \ref{com}, where there indeed exist a broad structure around $D\bar{D}^*$ threshold and its reflection. To some extent, this fact provides a direct test to our prediction presented here.

\vfil

\section*{Acknowledgment}

X.L. would like to thank Chang-Zheng Yuan for suggestive discussion. This project is supported by the
National Natural Science Foundation of China under Grants Nos. 10705001,
No. 11005129, No. 11035006,
No. 11047606, the Ministry of Education of China (FANEDD under Grant
No. 200924, DPFIHE under Grant No. 20090211120029, NCET under
Grant No. NCET-10-0442, the Fundamental Research Funds for the
Central Universities), and the West Doctoral Project of Chinese
Academy of Sciences.

\end{document}